\documentclass[review,fleqn,sort&compress]{elsarticle}
\usepackage{amssymb,amsbsy,amsthm,amsmath,amsfonts,amssymb,amscd}
\usepackage{geometry}
\usepackage{babel,blindtext}
\usepackage{bm} 

\usepackage{graphicx}
\usepackage{float}
\usepackage{multirow}
\usepackage{subfigure}
\usepackage{lineno}
\usepackage{hyperref}
\hypersetup{colorlinks, citecolor=black, filecolor=black, linkcolor=black, urlcolor=black}
\usepackage{booktabs}
\usepackage[ruled,linesnumbered]{algorithm2e}
\usepackage{appendix}
\usepackage{algpseudocode}
\usepackage{mathtools,nccmath}
\usepackage{adjustbox}
\usepackage{cleveref}
\usepackage{dutchcal}
\usepackage[utf8]{inputenc}
\usepackage[scr=rsfso,cal=euler,frak=euler,bb=ams]{mathalfa}
\usepackage{longtable}
\usepackage{comment}
\usepackage{color}
\usepackage[dvipsnames]{xcolor}
\usepackage{cancel}
\usepackage{soul}
\colorlet{soulred}{red!40}

\usepackage{xparse}
\usepackage{mathrsfs}
\usepackage{upgreek}
\usepackage{colortbl}
\usepackage{hhline}
\newcolumntype{C}[1]{>{\centering\let\newline\\\arraybackslash\hspace{0pt}}m{#1}}

\begin{document}

\begin{frontmatter}
\title{A hybrid IFENN solver for generalizable modeling of phase-field fracture initiation and propagation}

\author{Panos Pantidis\corref{cor1}}
\author{Fouad Amin}
\author{Diab Abueidda}
\author{Mostafa E. Mobasher\corref{cor2}}
\cortext[cor1]{Corresponding author. \emph{E-mail address:} \texttt{pp2624@nyu.edu} (Panos Pantidis)}
\cortext[cor2]{Corresponding author. \emph{E-mail address:} \texttt{mostafa.mobasher@nyu.edu} (Mostafa Mobasher)}
\address{Civil and Urban Engineering Department, New York University Abu Dhabi, Abu Dhabi, P.O. Box 129188, UAE}

\begin{highlights}

\item IFENN models complete evolution of phase-field fracture.

\item The framework deploys a DeepOKAN for crack initiation and a CNN for propagation.

\item Novel boundary conditions or a weight-function ensure near-zero far-field values.

\item Training is purely physics-informed, without any labeled datasets.

\item The networks are trained only once and are applied on different geometries.

\end{highlights}

\begin{abstract}

In this paper we demonstrate how the Integrated Finite Element Neural Network (IFENN) framework can effectively model the entire evolution of phase-field fracture, including the initiation and propagation stage, across generalizable geometries. IFENN is a hybrid scheme for coupled computational mechanics problems, tightly coupling a standard FEM solver (mechanical equilibrium) with a pre-trained neural network (coupled field). In this work, the phase-field diffusion equation is approximated with: i) a DeepONet architecture with Kolmogorov-Arnold networks in the trunk and branch (DeepOKAN) for the initiation stage, and ii) a Convolution Neural Network (CNN) for the propagation stage. Both networks are trained only once, on a benchmark geometry, using a purely physics-informed approach based on the maximum strain energy and the phase-field variable. The training process utilizes an extremely small number of training increments and only a limited number of Gauss points that are strategically sampled from the fracture process zone. These features enable a substantial decrease of the offline training cost. To address the extrapolation of the DeepOKAN predictions in regions away from the crack tip during the inference stage, we implement a set of artificial boundary conditions to enforce the near-zero values in the far-field predictions. We showcase the flexibility and numerical accuracy of the proposed methodology across both the training and unseen geometries.

\end{abstract}

\begin{keyword}
\texttt IFENN \sep phase-field fracture \sep initiation \sep hybrid modeling  
\end{keyword}

\end{frontmatter}


\newpage
\section{Introduction}
\label{Section:Introduction}

\subsection{Literature review}

The simulation of complex, real-world mechanics problems requires capturing highly non-linear, coupled interactions across multiple physical fields. Traditional numerical frameworks, such as the Finite Element Method (FEM), remain the primary pathway of engineering computational analysis, building their rigor over decades of continuous research and development \cite{reddy2026introduction, hughes2003finite}. However, when applied to large-scale coupled problems, these methods still face significant challenges, typically associated with very intense computational costs and memory resources \cite{kennedy2014parallel}. While several remedies have been proposed to alleviate such bottlenecks, including enhanced solution algorithms \cite{kristensen2020phase}, HPC-deployment with parallel computing \cite{neiva2019scalable}, domain decomposition methods \cite{farhat1991method}, and adaptive re-meshing techniques \cite{cornejo2020combination}, advancing the efficiency of numerical solvers remains a critical hurdle in computational mechanics.

Fracture modeling is a distinct example of such a computationally demanding problem. Among the various methods which can be used to model fracture, the phase-field (PFF) approach has attracted significant attention over the past decades \cite{ambati2015review}. Established as a variational energy minimization framework \cite{francfort1998revisiting}, PFF regularizes the discrete crack into a continuous, smeared representation, through an auxiliary kinematic parameter $\phi$. By avoiding explicit crack tracking, the method can capture complex phenomena such as crack branching or coalescence \cite{zhou2019propagation, spatschek2011phase}. However, it is also associated with a notoriously high computational overhead, which stems from the need for very fine discretization in the vicinity of the open crack-tip and along the crack trajectory. This expenses become even more dominant in the case of 3D, large-scale, multiple-cracked domains, and to this date, the acceleration of phase-field fracture simulations remains an actively explored research topic \cite{svolos2022fourth, storvik2021accelerated, yang2023acceleration}. 

Machine-learning (ML) methods have recently emerged as a promising alternative with tangible, practical potential. While ML-based methods have found extensive applications in other problems in solid mechanics, such as anisotropic elasticity  \cite{zhang2024machine}, hyperelasticity \cite{vlassis2020geometric} and plasticity \cite{huang2020machine}, their application in purely PFF problems still remains rather elusive. In the context of PFF, ML-based approaches have primarily manifested as \textit{simulation substitution frameworks}. These frameworks aim to completely bypass conventional numerical solvers, by utilizing neural network surrogates. Several studies have followed this modeling paradigm \cite{ghaffari2023deep, kiyani2025predicting, goswami2020transfer, goswami2022physics}, with the common objective of predicting the evolution of the displacement field and phase-field profiles under arbitrary loading conditions. However, several inherent factors can impede the computational efficiency of such methods, including the history-dependent nature of the phenomenon, the need for predicting multiple outputs with very small numerical nuances, as well as the need for generalization of these models in unseen geometries. Eventually, all of these constraints lead to either complete or partial re-training for different scenarios \cite{goswami2020transfer, manav2024phase, zheng2022physics}, which ultimately yield prohibitively long training time and hinder the adoption of such models for real-time problems. Beyond pure simulation substitution, frameworks focusing on embedding sufficient physics into the network design were proposed more recently, such as in \cite{dammass2025neural} and \cite{aldakheel2025physics}. Nevertheless, there remains a critical need to develop robust, physics-guided, and computationally efficient methods which are scalable to real-world problems.


\subsection{A hybrid paradigm: Integrated Finite Element Neural Network (IFENN)}

The Integrated Finite Element Neural Network (IFENN) framework is a hybrid modeling approach developed for coupled problems in solid mechanics. IFENN employs a staggered solution scheme with two distinct solvers: a classical FEM-based solver handles the mechanical equilibrium and computes the displacement field, including residual minimization, while a pre-trained neural network (NN) acts as the second solver and approximates the solution of the coupled governing PDE. The key feature is that the two platforms exchange information interchangeably, at every load increment, thus avoiding blind network predictions and error accumulation over long time periods. This hybrid approach achieves a twofold objective: i) it significantly accelerates the total solution time by leveraging the rapid online inference of the network compared to computationally intensive FEM, and ii) it preserves solution accuracy by obtaining the displacement field through residual minimization of conventional FEM. To date, IFENN has been successfully implemented in the modeling of poromechanics formulations \cite{amin2026fenn}, coupled thermoelasticity problems \cite{amin2026fenn, abueidda2024variational, abueidda2024fenn} and non-local gradient damage \cite{pantidis2023integrated, pantidis2023116160, pantidis2024fenn}. More recently it was implemented for phase-field fracture modeling \cite{pantidis2026integrated}, but we emphasize that this study was concerned only with the propagation stage of the crack. The initiation stage, which constitutes a significant and even more challenging-to-capture portion of the analysis, was still handled by a typical FEM solver. Therefore, capturing the full fracture evolution remains an open challenge for such tightly-coupled hybrid solution schemes.

\subsection{Scope and Outline}
\label{Scope_and_Outline}

The main objective of this study is to bridge the aforementioned gap and implement IFENN on the entire evolution of phase-field fracture, including both the initiation and propagation stage. In the design of this study, a key priority is to preserve already achieved flexibilities of IFENN from previous works, including the purely physics-based training \cite{pantidis2026integrated, pantidis2023116160, pantidis2023integrated}, as well as the seamless generalization of the same network across different geometries \cite{pantidis2026integrated}, mesh discretizations \cite{pantidis2024fenn}, and loading/boundary conditions \cite{amin2026fenn}. To this end, the main novel contributions of this work are summarized below:

\begin{itemize}

\item For the first time, we showcase the feasibility of IFENN to model the initiation stage of phase-field fracture. We highlight that this is an outstandingly challenging problem, due to its extreme localization and exponential, highly non-linear growth. The neural network component of IFENN for this stage is a deep neural operator with Kolmogorov-Arnold networks (DeepOKAN) in the trunk and branch.

\item We propose two approaches to enforce near-zero far-field predictions for the initiation stage: i) a set of novel artificial boundary conditions, which enables the DeepOKAN to achieve excellent extrapolation beyond its crack-tip training zone, and ii) a local weight-function approach, which effectively vanishes phase-field values in the far-field. Also, by strategically sampling the DeepOKAN sensors within the fracture process zone, we achieve a balanced trade-off between training cost and inference accuracy.

\item Training of both DeepOKAN (initiation) and CNN (propagation) occurs on a strictly physics-informed basis, without any reliance on labeled datasets. The proposed networks are trained just once, on a benchmark single-crack geometry, and the same networks are effectively used on other geometries with differences in terms of finite element discretization and location of initial cracks. 

\item The networks are trained on an extremely small number of increments (8 for the DeepOKAN and 2 for the CNN), yet they deliver accurate inference over two orders of magnitude more increments. No causality constraints or sequence models are utilized for this purpose. 

\item This is the first time that IFENN is implemented with two different networks capturing sequential stages of the same problem. This is a significant departure from previous works in the field, which typically rely on a single network for the inference task. Our work showcases that the second network can still yield accurate predictions even when it is driven by predictions from another network, rather than by the 'ideal' FEM data.

\end{itemize}

The paper is structured as follows. Section \ref{Sec:Phasefield_theory} introduces the fundamental concepts of phase-field fracture theory. Section \ref{Sec:Methodology} presents the core of the proposed methodology, including the networks setup and training details, deployment within IFENN, and generalization to different geometries. Section \ref{Sec:One_time_training} discusses the one-time training of the networks, and Section \ref{Sec:Numerical_results} presents the numerical examples of the study. Finally, a discussion on the limitations and conclusions are provided in Section \ref{Sec:Conclusions}.

\section{Phase-field fracture for quasi-brittle materials}
\label{Sec:Phasefield_theory}

\begin{figure}[H]
    \centering
    \includegraphics[width=0.5\textwidth]{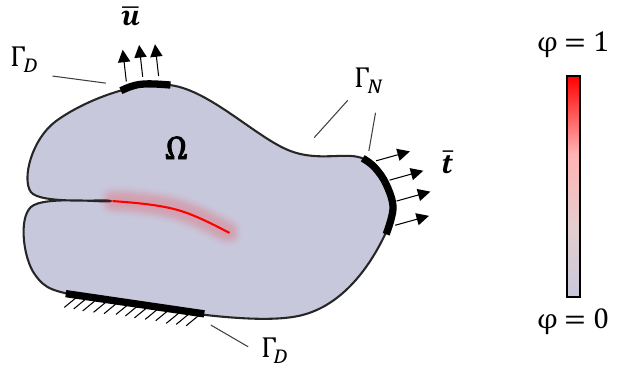}
    \caption{Schematic of a cracked domain with a sample phase-field damage contour}
    \label{Fig:Figure_Domain_Schematic}
\end{figure}

Let us consider the solid and deformable body $\Omega$ shown in Fig. \ref{Fig:Figure_Domain_Schematic}. The domain boundary $\partial \Omega$ is split in two subsets, $\Gamma_{D}$ and $\Gamma_{N}$, with Dirichlet and Neumann boundary conditions imposed respectively. Adopting the small deformation theory, the strain tensor is defined as: 

\begin{equation}
\boldsymbol{\epsilon} = \frac{1}{2} [ \nabla \boldsymbol{u} + (\nabla \boldsymbol{u})^\mathrm{T}]
\label{Eqn:EnergyFunctional}
\end{equation}

\noindent where $\boldsymbol{u}(\boldsymbol{x}): \Omega \times \mathbb{R}^+ \rightarrow \mathbb{R}^\delta$ represents the displacement field, $\boldsymbol{x} \in \Omega$ is a material point, $\delta \in \{1,2,3\}$ denotes the spatial dimension, $\nabla$ is the gradient operator, and the superscript $\mathrm{T}$ is the tensor transpose.

The phase-field approach to fracture mechanics \cite{bourdin2000numerical, francfort1998revisiting} approximates the sharp crack topology with a continuous phase-field variable $\phi(\boldsymbol{x}): \Omega \times \mathbb{R}^+ \rightarrow [0,1]$, with $\phi = 0$ representing the intact material and $\phi = 1$ denoting the fully cracked state. According to the AT2 model \cite{ambrosio1990approximation}, the deformation of the body and the crack topology are determined by minimizing the following total energy functional (neglecting body forces):

\begin{equation} 
    \mathcal{E}(\boldsymbol{u}, \phi) = \int_\Omega \Psi(\boldsymbol{\varepsilon}, \phi) d\boldsymbol{x} + \int_\Omega \frac{G_c}{2 l_c} \left( \phi^2 + l_c^2 \Vert \nabla \phi \Vert^2 \right) d\boldsymbol{x} - \int_{\Gamma_N} \bar{\boldsymbol{t}} \cdot \boldsymbol{u} d\boldsymbol{s} \, ,
\label{Eqn:EnergyFunctional}
\end{equation}

\noindent where $\Psi$ is the degraded strain energy density due to damage, $G_c$ is the critical energy release rate, $l_{c}$ is the characteristic length scale, and $\bar{\boldsymbol{t}}$ is the traction vector applied on $\Gamma_N$. 

The extent of material damage is controlled by the degradation function $g(\phi)$, which in this study is selected as $g(\phi) = (1-\phi)^2$. Minimizing the total energy functional in Eqn. \ref{Eqn:EnergyFunctional} and employing the hybrid isotropic/anisotropic formulation proposed by \cite{ambati2015review}, yields the following set of coupled governing equations:

\begin{equation}
\left.
\begin{aligned}
\boldsymbol{\nabla} \cdot \boldsymbol{\sigma} & = 0 \\
\frac{G_c}{l_c} \phi - G_c \, l_c \Delta \phi + 2(1-\phi) H  & = 0  \\ 
\end{aligned} \right\} \, .
\label{Eqn:governing_PDEs}
\end{equation}

\noindent where $\boldsymbol{\nabla}$ is the divergence operator, $\Delta$ is the Laplace operator, and $\boldsymbol{\sigma}$ denotes the Cauchy stress tensor which is defined as: 

\begin{equation}
\boldsymbol{\sigma} = (1-\phi)^2 \left[ \lambda \operatorname{tr}(\boldsymbol{\varepsilon}) \boldsymbol{I} + 2 \mu \boldsymbol{\varepsilon} \right]
\label{Eqn:StressCauchy}
\end{equation}

\noindent where $\lambda$ and $\mu$ are the first and second Lam\'e constants respectively, $\operatorname{tr}$ is the trace operator, $\boldsymbol{I}$ is the second-order identity tensor, and $H$ is the maximum strain energy density variable (also called history variable), defined as:

\begin{equation}
    H(\bm{x}) = \max_{t=t_0, \dots, t_{n-1}} \Psi_0^+ (\bm{x}, t) \, \quad \text{in} \quad \bm{x} \in \Omega ,
\label{Eqn:HistoryVar}
\end{equation}

The mathematical formulation of the problem is completed by the following set of boundary conditions:

\begin{equation} \left.
\begin{aligned}
    \boldsymbol{u}(\boldsymbol{x}) = \bar{\boldsymbol{u}}(\boldsymbol{x}) \quad \text{on} \quad \boldsymbol{x} \in \Gamma_D \\
    \boldsymbol{\sigma} \boldsymbol{n}(\boldsymbol{x})= \bar{\boldsymbol{t}}=\boldsymbol{0} \quad \text{on} \quad \boldsymbol{x} \in \Gamma_N \\
    \nabla \phi \cdot \boldsymbol{n}(\boldsymbol{x}) = 0 \quad \text{on} \quad  \boldsymbol{x} \in \partial \Omega
\end{aligned} \right\} \, ,
\label{Eqn:BCs}
\end{equation}

\noindent where $\bm{n}(\bm{x})$ is the unit outward vector normal to the boundary at the point $\bm{x} \in \partial \Omega$.

\section{Methodology}
\label{Sec:Methodology}

This section outlines the key components of our methodology. First, Section \ref{Sec:Method_DeepOKANs} introduces the network for the initiation stage of phase-field fracture, which is a physics-informed operator with Kolmogorov-Arnold networks (PI-DeepOKAN). Section \ref{Sec:Method_PICNN} describes the setup of the physics-informed convolutional neural network (PI-CNN) which is used to capture the propagation stage of phase-field. Finally, Section \ref{Sec:Method_IFENN} presents the integrated IFENN workflow and its numerical implementation details.

\subsection{Physics-informed DeepOKAN for phase-field fracture initiation}
\label{Sec:Method_DeepOKANs}

\subsubsection{Why DeepOKAN?}
\label{Sec:Method_Why_DeepOKANs}

The DeepONet architecture is a neural network-based approximation of an operator $(\mathcal{G})$ \cite{lu2021learning}. Unlike traditional neural networks, which map input variables to output variables, DeepONet maps an input function to an output function. The DeepONet employs a dual-network architecture, which are termed \textit{branch} and \textit{trunk}. The branch network encodes the discretized input function at specific 'sensor' locations, while the trunk encodes the locations where the output function is evaluated. In its vanilla form, as proposed in \cite{lu2021learning}, both branch and trunk employed Multi-Layer Perceptrons (MLPs). However, a key strength of the DeepONet architecture is its flexibility to accommodate different types of networks beyond simple MLPs, such as GRUs \cite{amin2026fenn}, LSTMs \cite{he2024sequential} and CNNs \cite{huang2024porous}. 

Abueidda et al. introduced DeepOKAN \cite{abueidda2025deepokan}, a DeepONet variant based on Kolmogorov-Arnold networks (KANs) \cite{liu2025kan}. In the latter architecture, the fixed linear weights of conventional MLPs are replaced with learnable activation functions at the edges, such as B-splines or Radial Basis Functions (RBFs) \cite{somvanshi2025survey}. While the comparison between DeepOKANs and traditional DeepONets remains an active research topic across several areas, prior work in solid mechanics problems \cite{abueidda2025deepokan} suggested that the KAN choice offered a comparative advantage against MLPs for similar levels of network complexity (as measured by the total number of trainable parameters). At the same time, several studies in the literature report significant issues when training MLPs for phase-field fracture, such as parameter-sensitivity and convergence fragility \cite{manav2024phase}. To this end, these challenges motivate the adoption of the RBF-DeepOKAN proposed in \cite{abueidda2025deepokan} and examine its expressivity and robustness for phase-field fracture problems. 

Before we discuss the technical details of the developed network, we highlight a few additional motivations for adopting an operator-based network. First, neural operators show significant generalization potential, and the study of \cite{amin2026fenn} demonstrated this feature for different loading scenarios within the IFENN context. In this study, we extend this even further: we train only once, for a benchmark single-crack geometry, and then the same DeepOKAN is deployed to predict initiation across diverse geometries, without further architectural modifications. The details of this capability are discussed in Section \ref{Sec:Method_crackgeneralization}. Second, though this is not investigated here and it is left for future work, these architectures offer seamless scalability from 2D to 3D domains. This extension is achieved by simply augmenting the trunk network’s input with the $z$-coordinate. Although a new network must be re-trained for the 3D setup, the underlying principles and overall architecture remain identical, while the number of trainable parameters does not increase prohibitively.

\subsubsection{DeepOKAN architecture}
\label{Sec:Method_DeepOKAN_architecture}

\begin{figure}[H]
    \centering
    \includegraphics[width=0.95\textwidth]{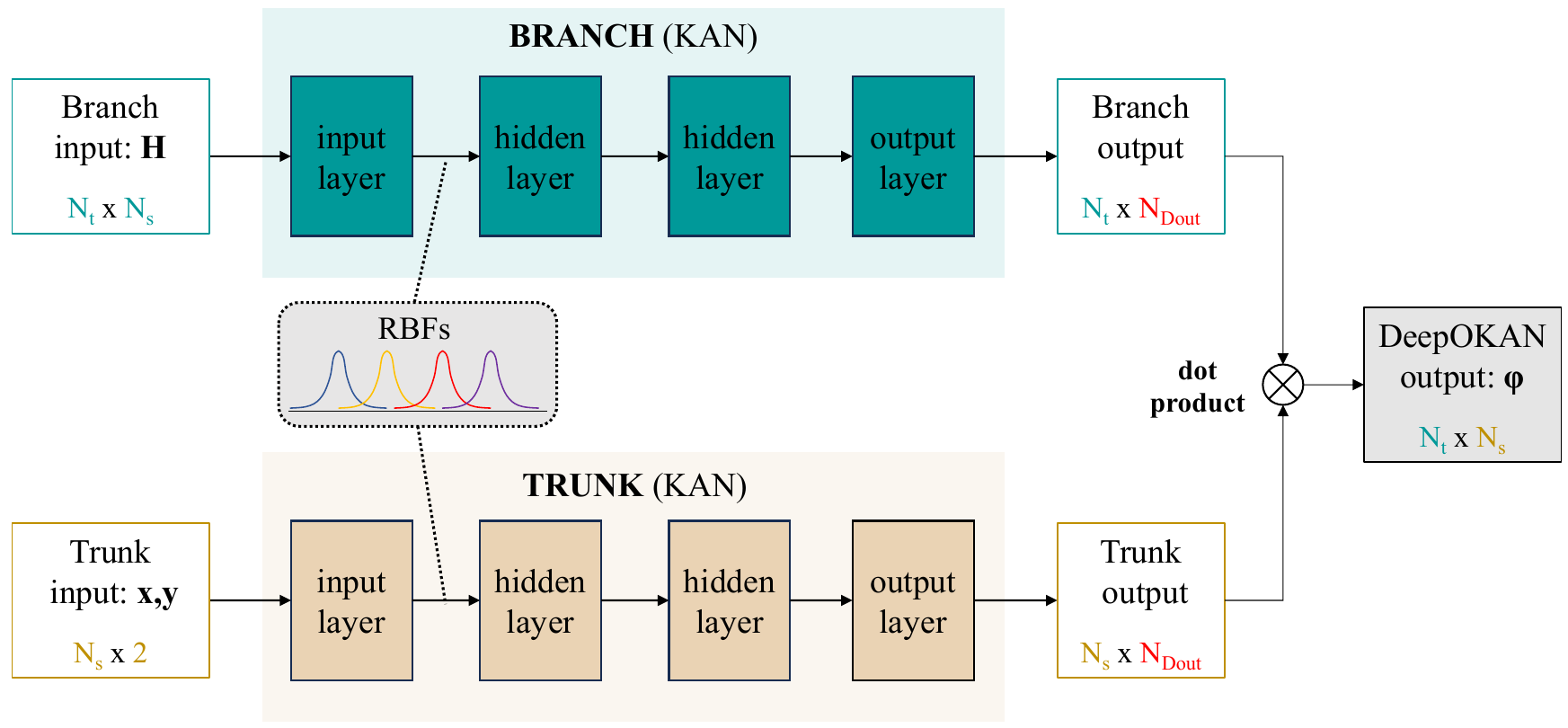}
    \caption{Architecture of the adopted DeepOKAN, with Radial Basis Functions (RBFs) as the activation functions for all the edge connections.}
    \label{Fig:Figure_DeepOKAN_Schematic}
\end{figure}

The architecture of the adopted RBF-DeepOKAN is shown in Fig. \ref{Fig:Figure_DeepOKAN_Schematic}. The branch network encodes the $H$ profile at selected $N_{s}$ sensor locations across different $N_{t}$ time increments. Accordingly, the size of the training input matrix for the branch is $[N_{t} \times N_{s}]$. The trunk network encodes the 2D coordinates of these sensors, receiving an input matrix of $[N_{s} \times 2]$. Each hidden layer of both the branch and the trunk has a width of $N_{h}$, and all layers have $N_{g}$ number of learnable RBFs. The common output dimension of the two sub-networks is $N_{Dout}$, yielding a final output of size $[N_{t} \times N_{Dout}]$ and $[N_{s} \times N_{Dout}]$ for the branch and the trunk respectively. Using the dot-product operation, the final shape of the network output is $[N_{t} \times N_{s}]$, and contains the predicted phase-field values across the selected time increments at the specific sensor locations. 

The selected architectural configuration is justified by several key considerations which are discussed below. For the branch, by tying $N_{s}$ to the network architecture and leaving the temporal dimension as independent, we obtain a specific trade-off. This choice restricts the number of sensors to remain the same across all future evaluations of the trained network, for example across a new geometry. Therefore, these sensors need to be carefully selected in order to maximize their representation capacity, a process which we detail in Section \ref{Sec:Method_DeepOKAN_sampling}. At the same time, this choice allows the model to be trained on an arbitrary number of time increments and enables inference at any individual time step, a flexibility which is essential for the online stage of IFENN. For the trunk, we note that by treating the spatial dimension $N_{s}$ as the independent one, the network can receive an arbitrary number of query points in the online stage, enabling the use of the same network across varying mesh densities. 

\subsubsection{Sampling of collocation points (sensors)}
\label{Sec:Method_DeepOKAN_sampling}

The phase-field crack initiation is an extremely localized phenomenon, which evolves mostly at the close proximity of the crack-tip. To achieve satisfactory solution accuracy, this region is typically discretized with a very fine finite element mesh. However, utilizing all Gauss points within the domain as network sensors is computationally infeasible, as it would result in an unreasonably high branch input dimension (as discussed in Section \ref{Sec:Method_DeepOKAN_architecture}). Therefore, to maximize training efficiency while retaining adequate model accuracy, a strategic sampling process for the sensors needs to be employed. In this study we propose two sampling strategies, each tied to our approach for training and generalizing to other geometries, which are detailed below. \vspace{0.4em}

\begin{figure}[b!]
    \centering
    \includegraphics[width=1\textwidth]{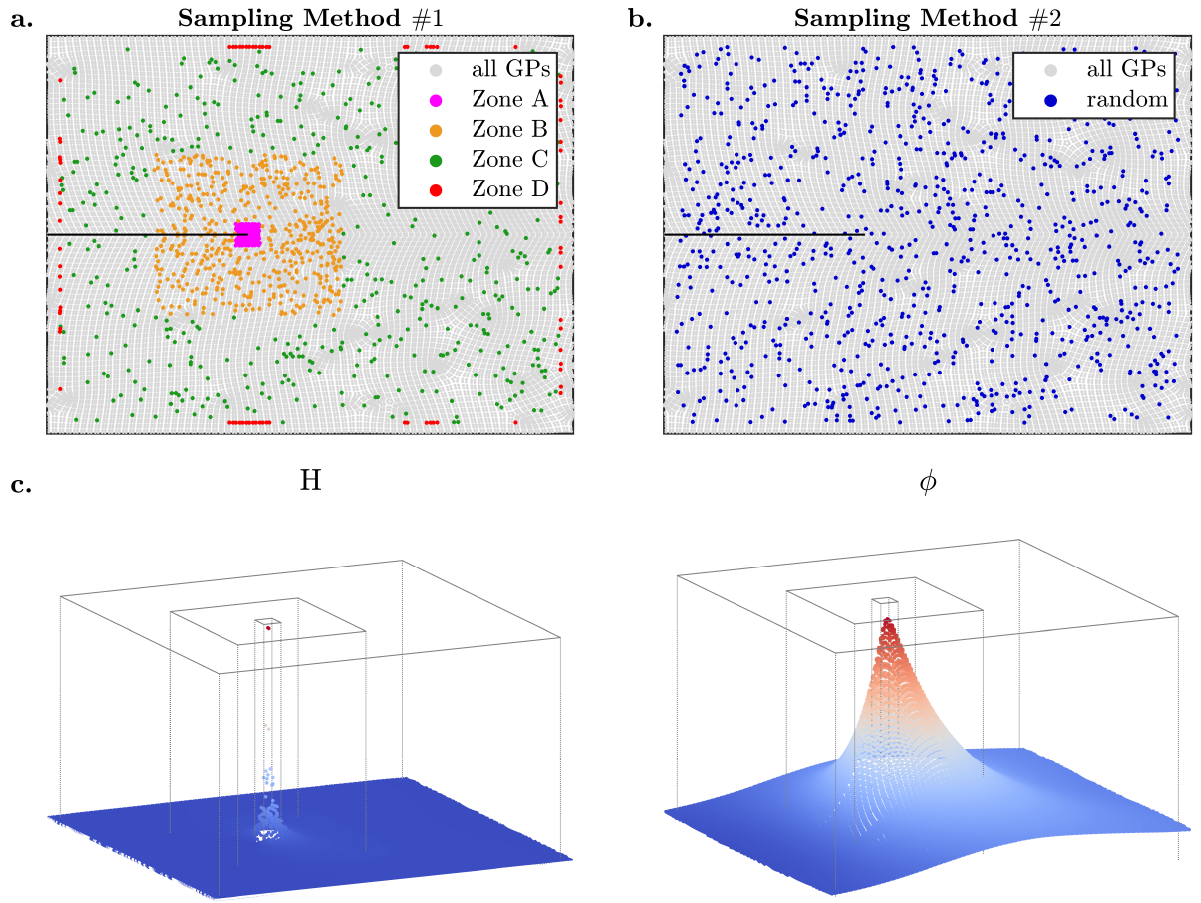}
    \caption{{\bf{a.}} First sampling method (SM1), where colored Gauss points indicate sensor locations across zones A–D. {\bf{b.}} Second sampling method (SM2), featuring uniformly and randomly selected Gauss points. {\bf{c.}} Representative profiles of $H$ and $\phi$, illustrating the information contained within each zone.}
    \label{Fig:Figure_SMs}
\end{figure} 

\noindent {\underline{Sampling Method $\#1$ (SM1)}}

As shown in Fig. \ref{Fig:Figure_SMs}a, this approach utilizes the following spatial discretization:

\begin{itemize}

    \item Zone A: This is a square region spanning the length of 6 finite elements in the $x-$ and $y-$direction. These elements contain the most meaningful information of the history variable, since $H$ exhibits a rapid drop outside this area. All the Gauss points of these elements are utilized as sensors, and they are denoted as $N_{GP,A}$. 

    \item Zone B: This region comprises a square domain extending four characteristic lengths in both the $x$- and $y$-directions. In this zone, while the history variable $H$ has decayed significantly, the phase-field variable $\phi$ remains non-negligible relative to the rest of the domain. A fraction of the constituent Gauss points are randomly selected as sensors to represent this region, denoted as $N_{GP,B}$.
    
    \item Zone C: This is a rectangular domain bounded by the distance at which $\phi$ becomes essentially negligible. Similar to Zone B, the contribution of this area is represented by a fraction of its Gauss points that is randomly selected and denoted as $N_{GP,C}$.
    
    \item Zone D: This is the ``interface'' of Zone C with the rest of the domain. Since this is not an actual, physical boundary, we select a sparse set of Gauss points that are located as close as possible to this virtual interface, acting as peripheral sensors. This set of sensors is denoted as $N_{GP,D}$, and we note that it is equally distributed across the four sides of zone D.

\end{itemize}

Figure \ref{Fig:Figure_SMs}c displays representative profiles of $H$ and $\phi$, where the projected boundaries delineate each partitioned region. Overall, SM1 ensures that each zone provides meaningful information to the network, while maintaining a reasonably low number of training sensors. \vspace{0.4em}

\noindent {\underline{Sampling Method $\#2$ (SM2)}}

Unlike the zoned partitioning of the previous approach, Sampling Method 2 (SM2) treats the entire active domain (Zone C) as a single region. Sensors are selected via uniform random sampling across all available Gauss points, as shown in Fig. \ref{Fig:Figure_SMs}b, and they are completely independent of local field gradients. The set of the selected Gauss points in SM2 is denoted as $N_{GP}$. Although SM2 does not explicitly prioritize the crack-tip core, it offers two practical benefits. First, it eliminates the arbitrariness and algorithmic overhead of defining the sub-zones. Second, it serves as a benchmark baseline to evaluate whether the neural network can learn localized features from globally sampled data, and offers a comparative basis against the more sophisticated SM1.

\subsubsection{DeepOKAN training setup}
\label{Sec:Method_DeepOKAN_training}

The learning objective of the DeepOKAN is to capture the physical relationship between the history variable $H$ and the phase-field variable $\phi$. In this study we completely omit labeled data from the training process, adopting a pure physics-informed network variant. The input dataset includes the $x-$ and $y-$coordinates and $H$ profiles of the sensors. This information is acquired from a single analysis on a benchmark single-notch geometry. We note that a typical staggered FEM analysis of phase-field fracture requires several hundreds of increments. However, for the network training, we select the $H$ profile at only a very small number of increments, relying on the network interpolation capability for the online IFENN stage. In this study we propose two training strategies, which are discussed below. \vspace{0.4em}

\noindent {\underline{Training strategy $\#1$ (TS1)}}

In the first method, henceforth termed $TS1$, the network is trained on two objectives. First, it learns to minimize the strong-form residual of the governing PDE across the entire sensor ensemble (zones A-D). This ensures that the network captures the $H-\phi$ relationship in the area where both profiles are non-negligible. Next, in order to construct a network that predicts constant, near-zero $\phi$ values in the rest of the domain (outside of zone D), a set of ``vanishing'' boundary conditions is enforced on the sensors of zone D. These are \textit{artificial} constraints which teach the network to satisfy both Dirichlet-like and Neumann-like conditions on these sensors, minimizing the magnitudes of both $\phi$ and its gradient $\nabla \phi$ respectively. These conditions are motivated by the observation that both the phase-field and its associated spatial flux effectively vanish at the periphery of the localized initiation zone. Also, we underline that this modification is partially inspired by the non-local gradient damage formulation of \cite{peerlings1998gradient} and its accompanying Neumann-type boundary condition. 

With these remarks in mind, the loss function of the network reads as:

\begin{equation}
    \mathcal{L}_{total,1} = \mathcal{L}_{PDE} + \mathcal{L}_{BCs}
\label{Eqn:Loss1}
\end{equation}

\noindent where:

\begin{equation}
    \mathcal{L}_{PDE} = \underbrace{\bigg\| \frac{G_c}{l_c} \phi(x,y) - G_c l_c \nabla^2 \phi(x,y) - 2\big(1 - \phi(x,y)\big) H(x,y) \bigg\|_2}_{\text{Interior Residual (Zones A--D)}}
\label{Eqn:Loss_PDE}
\end{equation}

\begin{equation}
    \mathcal{L}_{BCs} = \underbrace{ \left\| \phi(x,y) \right\|_2 + \left\| \nabla^2 \phi(x,y) \right\|_2 }_{\text{Vanishing Constraints (Zone D)}}
\label{Eqn:Loss_BCs}
\end{equation}

\noindent and $\left\|. \right\|$ is the second norm of a vector. 

The network training is an optimization task defined as:

\begin{equation}
    \theta^* = \arg \min_{\phi} \mathcal{L}_{total,1}(\theta)
\label{Eqn:Loss_theta1}
\end{equation}

\noindent where $\theta$ are the network trainable weights, and $\theta^*$ is the optimized set of weights that minimizes the loss function $\mathcal{L}_{total,1}$. The final $\phi_{NN}$ profile reads as:

\begin{equation}
    \phi_{NN} = \mathcal{G}_{NN}(x,y,H ; \theta^*)
\label{Eqn:Loss_theta}
\end{equation}

As will be shown in the numerical examples, this method allows for near-zero predictions of $\phi$ at the GPs outside zone D, meaning the predicted phase-field profile effetively vanishes outside the training domain.

\noindent {\underline{Training strategy $\#2$ (TS2)}}

In the second approach, henceforth termed $TS2$, the network is trained solely on the governing PDE, minimizing the strong-form residual on the collocation points in zones A-D. In this case the loss function reads: 

\begin{equation}
    \mathcal{L}_{total,2} = \mathcal{L}_{PDE}
\label{Eqn:Loss2}
\end{equation}

\noindent where $\mathcal{L}_{PDE}$ is given in Eqn. \ref{Eqn:Loss_PDE}. If no further modifications are implemented on this setup, the resulting network will inherently struggle to resolve the vanishing values of $\phi$ in the far-field, given the complete absence of training data informing its behavior in that region. To circumvent this limitation, we introduce a localized weighting-function approach inspired by the window-function concept proposed in \cite{lai2026locally}. The authors of the latter work employ a dual-network architecture, superimposing a global smooth response with a localized sharp response via a coupling function. Here, we adopt a significantly more simplified strategy. We bypass the secondary network by directly introducing a spatial weight-function that is bounded between $0$ and $1$, which smoothly transitions to zero outside zone D. By evaluating this function on a GP-by-GP basis against the network's raw predictions, the far-field phase-field values are systematically suppressed. Consequently, this method allows for near-zero prediction of $\phi$ at the GPs outside zone D, yielding a physically reasonable phase-field profile. The proposed weight-function is formulated as:

\begin{equation}
w(x, y) = 
\begin{cases} 
\exp \left( -\left(\frac{r}{\sigma}\right)^p \right) & \text{if } r < \frac{R}{2} \\ 
0 & \text{otherwise} 
\end{cases}
\end{equation}

\noindent where the radial distance $r$ is computed relative to a shifted center $\mathbf{x}_c = [x_c, y_c]^T$ as:

\begin{equation}
r = \sqrt{(x - x_c)^2 + (y - y_c)^2}
\end{equation}

The shifted center coordinates are defined based on the crack tip position $\mathbf{x}_{\text{crack}} = [x_{\text{crack}}, y_{\text{crack}}]^T$ such that:

\begin{equation}
\begin{aligned}
x_c &= x_{\text{crack}} + 0.03 \\
y_c &= y_{\text{crack}}
\end{aligned}
\end{equation}

\begin{figure}[H]
    \centering
    \includegraphics[width=0.60\textwidth]{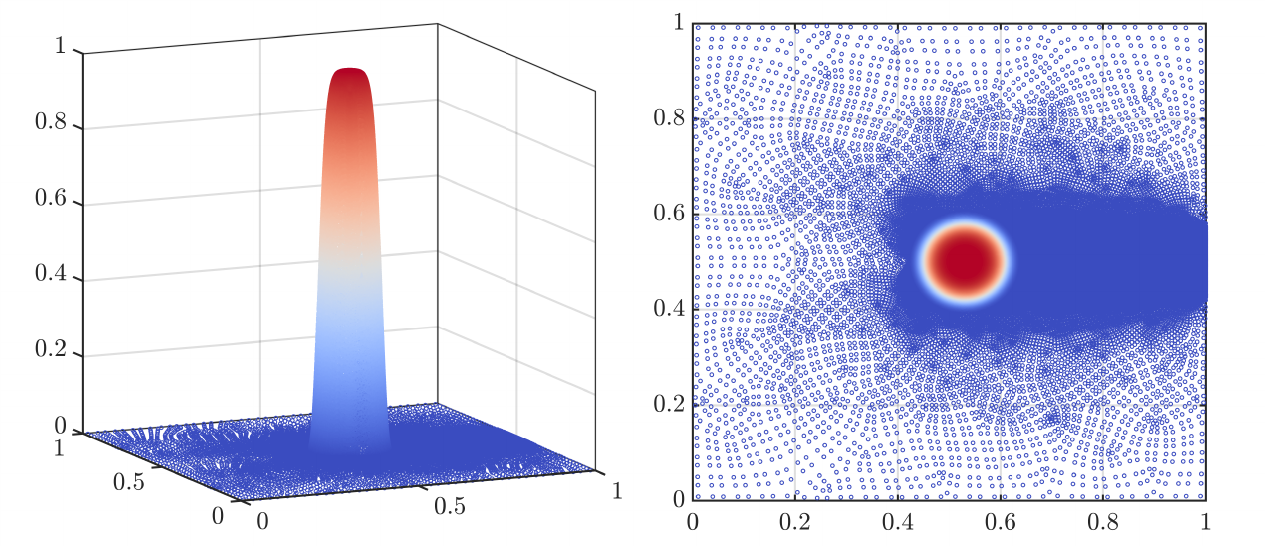}
    \caption{Schematic representation of the weight-function adopted for filtering the far-field $\phi$ values.}
    \label{Fig:Figure_g6_weight_function}
\end{figure}

The cutoff radius $R$ governing the domain of support is parameterized by:

\begin{equation}
R = \alpha_{l_c} \cdot l_{c}
\end{equation}

\noindent where $l_{c}$ is the characteristic length scale and $\alpha_{l_c}$ is a user-defined variable. A representative profile of this function is depicted in Fig. \ref{Fig:Figure_g6_weight_function}, using the parameter values that are detailed in Section \ref{Sec:One_time_training}.

Finally, we highlight two critical modifications compared to prior studies, including our previous work \cite{pantidis2026integrated}, that are implemented to enhance the network's expressivity. First, we omit bounding activation functions in the output layer. While bounding functions such as the \textit{sigmoid} or \textit{tanh} are used to constrain predictions within a specific range, for instance to bound $\phi$ between 0 and 1 during propagation, they are unsuitable for the initiation stage. During this stage, the phase-field variable must be free to attain any maximum value within the $[0, 1]$ interval, and therefore, such functions would create a bias towards a predefined upper bound, effectively impeding the learning objective. The second technical modification involves utilizing $log_{10}(H)$ as the branch input, instead of the raw variable $H$. The logarithmic transformation of the history variable creates a representation of the input space which is more numerically stable and better-suited for the network, leading to significant improvements in the training convergence and stability. To maintain the consistency of the physics-informed loss function, the history variable term in Eqn. \ref{Eqn:Loss_PDE} is reconstructed as $10^{H_{\text{input}}}$, where $H_{\text{input}}$ denotes the log-scaled sensor data.

\subsubsection{Generalization to different crack locations}
\label{Sec:Method_crackgeneralization}

\begin{figure}[H]
    \centering
    \includegraphics[width=1.0\textwidth]{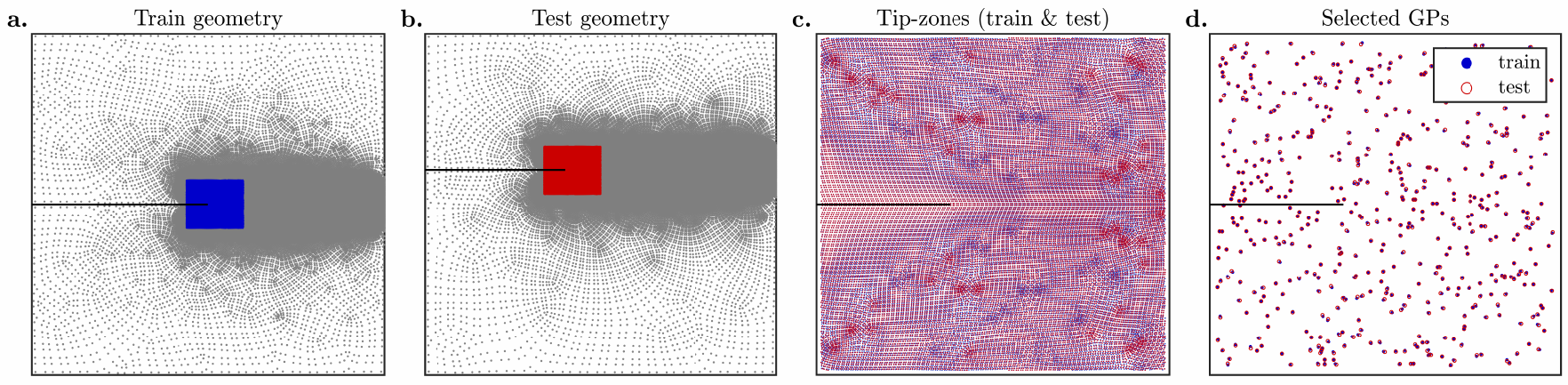}
    \caption{{\bf{a.}} Finite element mesh of a sample training geometry, with tip-zone highlighted in blue. {\bf{b.}} Finite element mesh of a sample testing geometry, with tip-zone highlighted in red. {\bf{c.}} Total number of Gauss points within the train and test tip-zones. {\bf{d.}} Sample train and test sensors (Gauss points), using the uniform random selection (SM2).}
    \label{Fig:Figure_Indices_ofOtherGeometries}
\end{figure}

Once the DeepOKAN is trained, its performance is strictly tied to the relative spatial order of the $H$ signals received by the branch network. Arbitrary sensor placement on new geometries will drastically degrade prediction accuracy, and therefore generalizing the trained network on new geometries requires that this sequential signal order is preserved. To satisfy this constraint, we establish a straightforward yet highly effective coordinate-matching scheme, that places the testing sensors (Gauss points) as close as possible to the original training coordinates. This approach is illustrated in Fig. \ref{Fig:Figure_Indices_ofOtherGeometries}, where the finite element meshes of the training and a testing geometry are displayed in Figs. \ref{Fig:Figure_Indices_ofOtherGeometries}a and b, respectively. Their corresponding crack-tip zones are highlighted in blue and red, and Fig. \ref{Fig:Figure_Indices_ofOtherGeometries}c displays the entire sets of Gauss points from both geometries. Utilizing the sampling strategy of SM2 as a representative case (though the same logic applies to SM1 as well), Figure~\ref{Fig:Figure_Indices_ofOtherGeometries}d maps the selected training sensors against their aligned testing counterparts. The spatial proximity between each matched pair is evident. As it will be shown in the numerical examples, this approach enables the DeepOKAN to readily predict the localized phase-field initiation on completely unseen geometries, without requiring re-training or architectural modifications.

\subsection{Physics-informed CNN for phase-field crack propagation}
\label{Sec:Method_PICNN}

The effectiveness of physics-informed convolution neural networks to capture the propagation stage of phase-field fracture within the IFENN framework has already been demonstrated in \cite{pantidis2026integrated}. For the sake of completeness, we provide below a concise description of this methodology, and the interested readers are referred to the previously cited work for further theoretical depth and technical details.

Fig. \ref{Fig:Figure_PICNN_Schematic} illustrates the general architecture and training setup of the PI-CNN. The model consists of a sequence of convolutional layers followed by the hyperbolic tangent function. The sigmoid function is applied after the final convolutional layer, bounding the final $\phi$ output between [0-1]. A key feature of the convolutional layers is the use of doubly-symmetric $5x5$ kernels, similar to \cite{pantidis2026integrated}, where only 6 out of the 25 parameters per kernel are independently trainable. CNNs necessitate a grid-based representation of both their input and output variables, therefore the unstructured $H$ are first interpolated onto a pixel-based grid, and subsequently the predicted $\phi$ values are mapped back to the original unstructured mesh. The PI-CNN receives two $H$ profiles which are sampled from the propagation stage of a benchmark FEM analysis. We note that $H$ is first normalized by the energy density scale $G_{c}/l_{c}$ and then passed as input into the network. Similarly, the network is trained on the dimensionless strong-form residual of the governing PDE:

\begin{figure}[t!]
    \centering
    \includegraphics[width=1.0\textwidth]{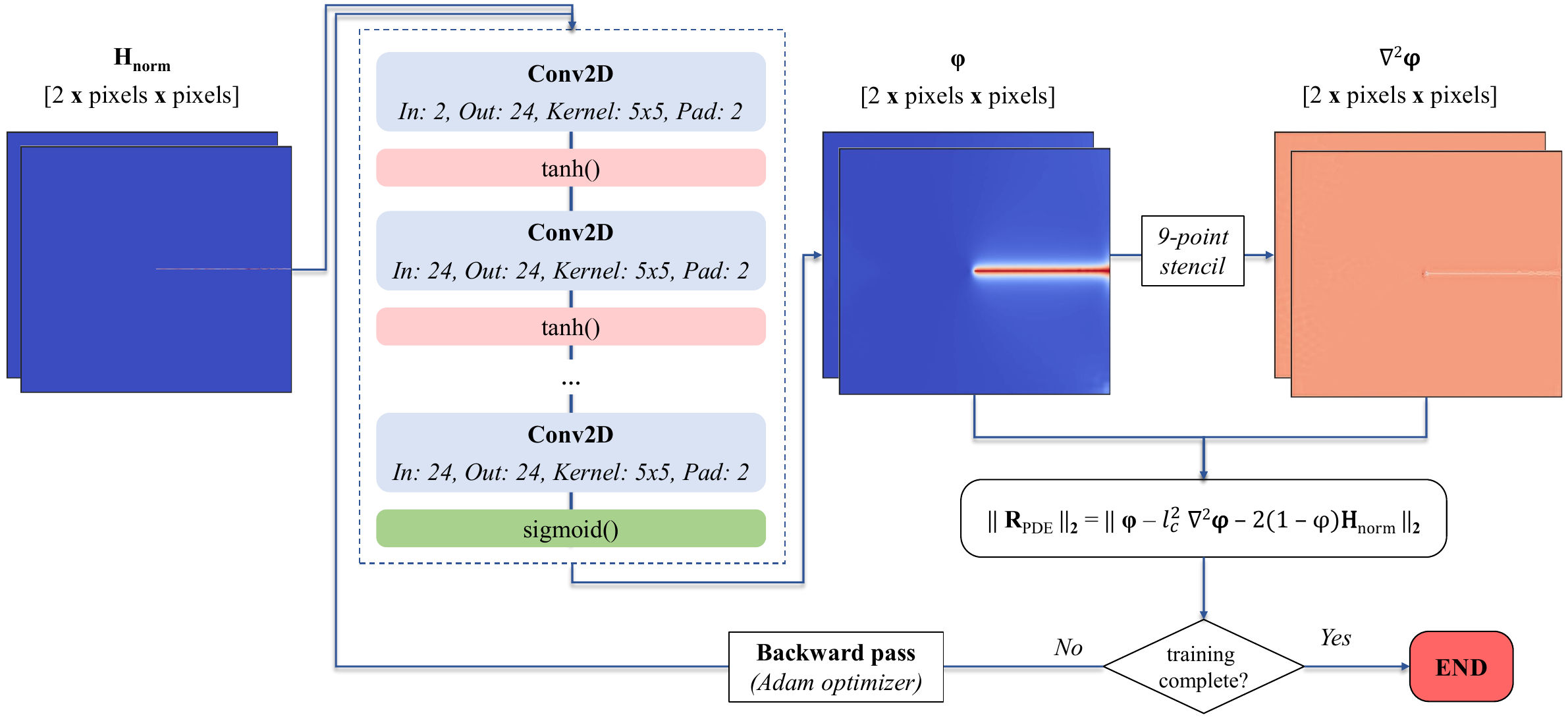}
    \caption{Illustration of the physics-informed CNN setup and training procedure.}
    \label{Fig:Figure_PICNN_Schematic}
\end{figure} 

\begin{equation}
    \mathcal{L}_{CNN} = \phi - l_{c}^{2} \nabla^2 \phi - 2(1 - \phi) \frac{H}{G_{c}/l_{c}}
\end{equation}

\noindent where the Laplacian operator $\nabla^2 \phi$ is computed using the 9-point stencil rule.

\subsection{IFENN for phase-field crack evolution}
\label{Sec:Method_IFENN}


The hybrid IFENN solver is comprised of three key stages. Since several detailed discussions of the IFENN implementation are available in the literature \cite{abueidda2024variational, abueidda2024fenn, pantidis2023integrated, pantidis2026integrated, pantidis2023116160, pantidis2024fenn, amin2026fenn}, we provide here only a brief summary of its main stages:

\begin{itemize}

    \item First, we perform a one-time, swift data-generation process. The goal is to reduce the offline computational cost of the data generation stage, which is typically substantial, and we achieve it either with a coarse mesh \cite{pantidis2024fenn} or with a coarse time-discretization FEM analysis \cite{pantidis2026integrated}.
    
    \item Second, a one-time training of a neural network. A key objective is to avoid re-training or transfer learning, which becomes a computational bottleneck for challenging problems such as phase-field fracture \cite{manav2024phase, goswami2020transfer}, and instead to design a flexible network that can be used in different scenarios, including different loading directions, geometries, and time-marching schemes.

    \item Third, the hybrid analysis of the target problem. This is the online stage where a conventional FEM solver computes the mechanical equilibrium (Eqn. \ref{Eqn:governing_PDEs}a), while the pre-trained network approximates the solution of the other PDE (in this work this is the phase-field diffusion, Eqn. \ref{Eqn:governing_PDEs}b).
    
\end{itemize}


Building upon the DeepONet-based IFENN framework \cite{amin2026fenn}, the numerical implementation is structured into three layers: a finite element method (FEM) layer, a neural network (NN) layer, and an operating system (OS) communication layer. The first layer resolves the mechanical equilibrium and is implemented in C++ using the open-source deal.II library \cite{arndt2023deal} and PETSc integration \cite{balay2019petsc} for parallel processing. It receives the phase-field profile from the previous time increment, solves Eqn. \ref{Eqn:governing_PDEs}a, and outputs the updated history variable profile. The second layer, written in PyTorch, approximates the phase-field equation expressed in Eqn. \ref{Eqn:governing_PDEs}b. This layer loads the pre-trained DeepOKAN and CNN models once, processes the $x, y, H$ inputs from the previous iteration, and predicts the new $\phi$ profile. A continuous data exchange between these two computing platforms is enabled by the OS communication layer. This layer leverages the operating system's shared memory and utilizes semaphore-based synchronization, preventing data corruption and coordinating the communication between the FEM and NN layers. 

\section{One-time training}
\label{Sec:One_time_training}

A primary source of the computational efficiency achieved with IFENN stems from the fact that the developed networks are trained only once on a benchmark geometry and subsequently reused across different cases. In this section we present the specific details of the developed networks and their one-time training.

\begin{figure}[H]
    \centering
    \includegraphics[width=0.8\textwidth]{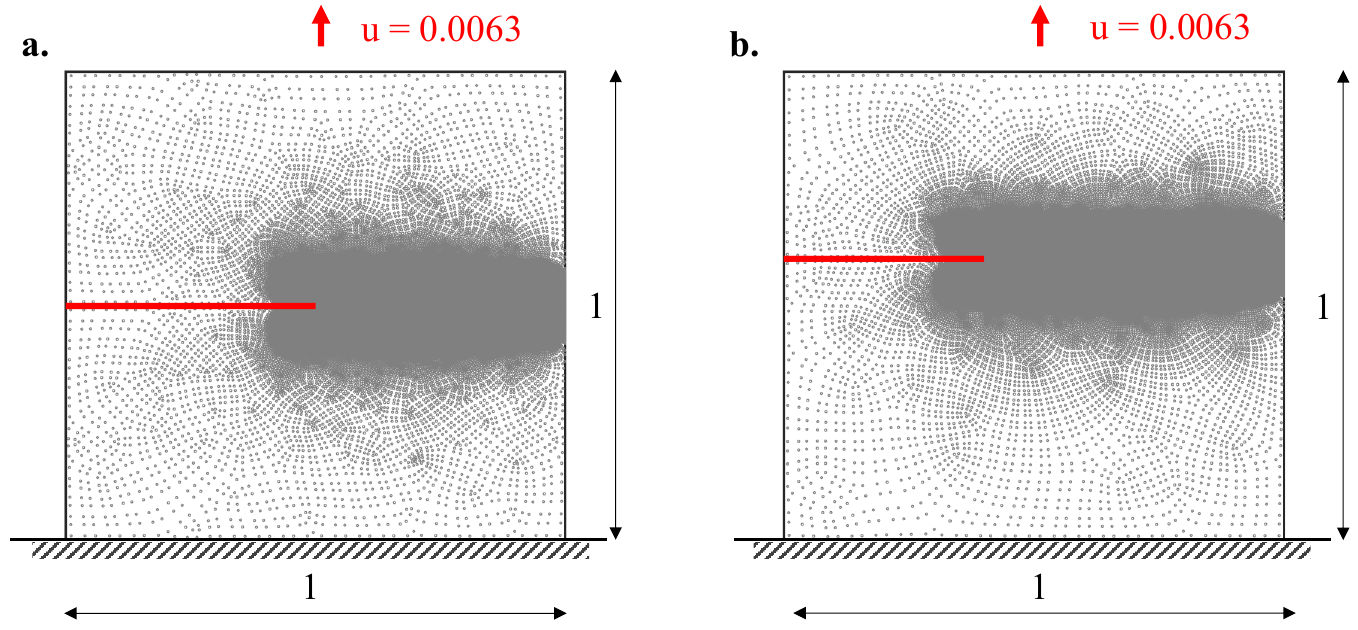}
    \caption{Schematic representations of {\bf{a.}} train geometry (Model A) and {\bf{b.}} test geometries (Model B).}
    \label{Fig:Figure_geometries}
\end{figure}

First, we perform a one-time, FEM analysis on a benchmark geometry, termed \textit{Model A}. This geometry is chosen as a square, symmetric, single-notch problem under tension. This case has been widely studied in the literature, and a schematic of the domain is shown in Fig. \ref{Fig:Figure_geometries}a. The bottom face is fixed in both directions, while a vertical upward displacement is applied on the top boundary. The Lame constants are $\lambda = 121154 N/mm^{2}$ and $\mu = 80770 N/mm^{2}$. The characteristic length scale is $l_{c} = 0.015 mm$, and the critical energy release rate is $G_{c} = 2.7 N/mm$. The AT2 model is adopted for the phase-field evolution \cite{ambrosio1990approximation}. A single-pass staggered FEM analysis is utilized, using initially a load increment of $\Delta u = 10^{-5} mm$ for the first 500 increments, followed by $\Delta u = 10^{-6} mm$ for the following 1000 increments. This setup closely mirrors established benchmarks in the literature, including the seminal work of Miehe et al \cite{miehe2010phase}. 

Once the FEM analysis is complete, we proceed with the one-time training of the networks. For the initiation stage we develop two variants of the DeepOKAN, one that is trained on TS1 and one on TS2. These are henceforth termed as DK-TS1 and DK-TS2. Both networks follow the general architecture of Fig. \ref{Fig:Figure_DeepOKAN_Schematic} and consist of two hidden layers. Also, both networks are trained on the $H$ profiles extracted from $N_{t} = 8$ distinct pseudo-increments: 100, 224, 309, 487, 1072, 1216, 1224, and 1232. We emphasize that the first training increment corresponds to an $H - \phi$ profile pair with maximum values on the order of $200$ and $0.01$, respectively, whereas the last training increment reaches maximum values of $4,000$ and $1.0$. This vast contrast of values, along with the extremely localized nature of their distribution, highlights the wide range of admissible values and numerical complexity that need to be captured by the networks. The structural hyperparameters and training configurations for both network variants are detailed below: 

\begin{itemize}
    \item $\textbf{DK-TS1 configuration}$: This variant is constructed with a hidden layer width of $N_{h} = 10$, number of grids $N_{g} = 25$, and an output dimension of $N_{Dout} = 512$. The network is optimized via the Adam optimizer for 5000 epochs at a learning rate of $1 \times 10^{-4}$. Training is executed using the SM1 approach, with $N_{GP,A} = 78$, $N_{GP,B} = 500$, $N_{GP,C} = 500$, and $N_{GP,D} = 80$, thus establishing an input layer width of $N_{s} = 1,158$.

    \item $\textbf{DK-TS2 configuration}$: This network utilizes a hidden layer width of $N_{h} = 14$, while maintaining $N_{g} = 25$ and $N_{Dout} = 512$. It is trained using the Adam optimizer for 10000 epochs with a learning rate of $5 \times 10^{-5}$. DK-TS2 is trained using the SM2 sampling strategy with uniformly distributed $N_{GP} = 1,000$ sensors.
    
\end{itemize}

\begin{figure}[H]
    \centering
    \includegraphics[width=1\textwidth]{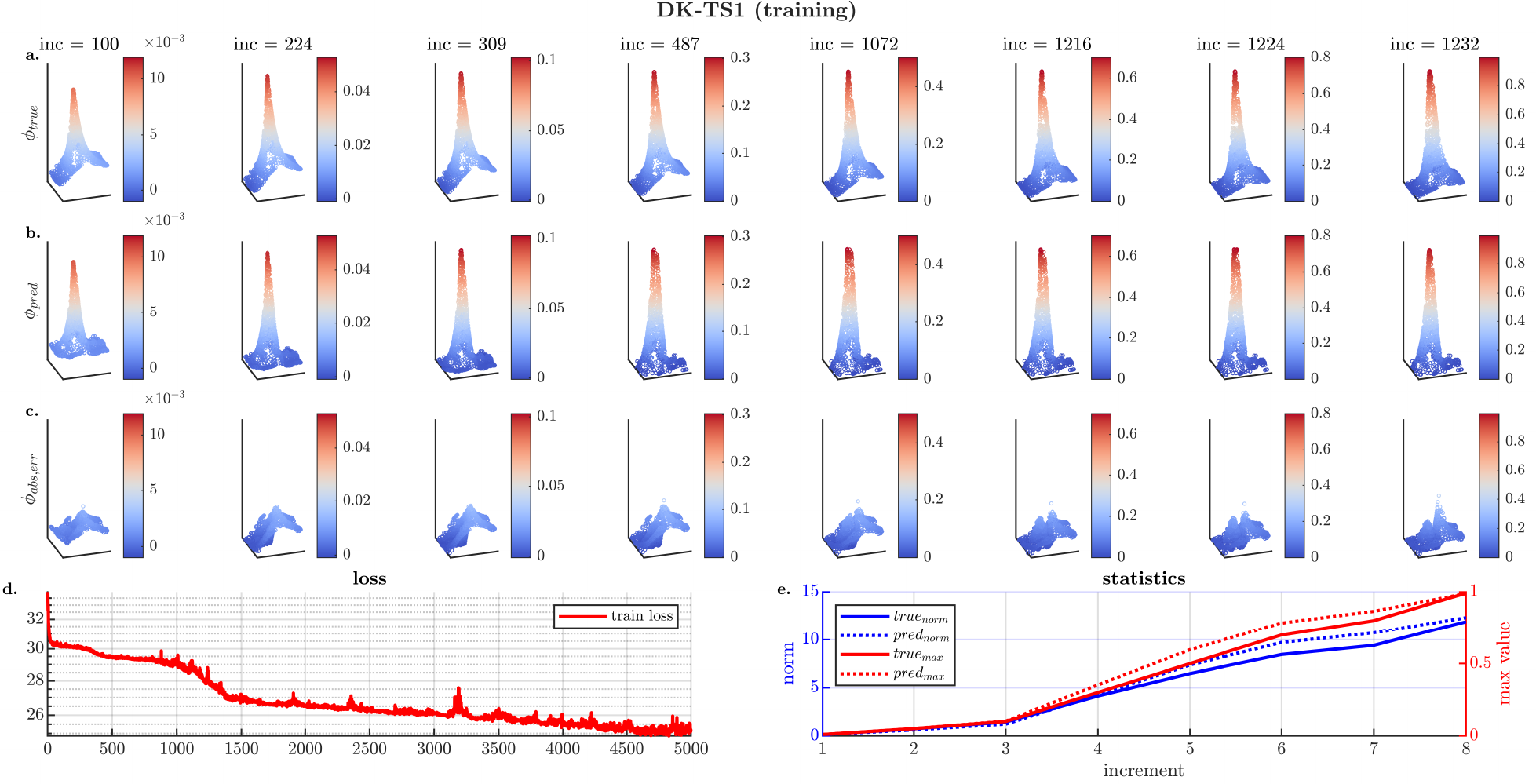}
    \caption{Training performance of DK-TS1. {\bf{a.}} True $\phi$ values. {\bf{b.}} Predicted $\phi$ values. {\bf{c.}} Absolute error $\phi$ values. {\bf{d.}} Evolution of training loss. {\bf{e.}} Metrics of $\phi$, including its $L^2$ norm and maximum value.}
    \label{Fig:Figure_DK_TS1_training}
\end{figure}

\begin{figure}[H]
    \centering
    \includegraphics[width=1\textwidth]{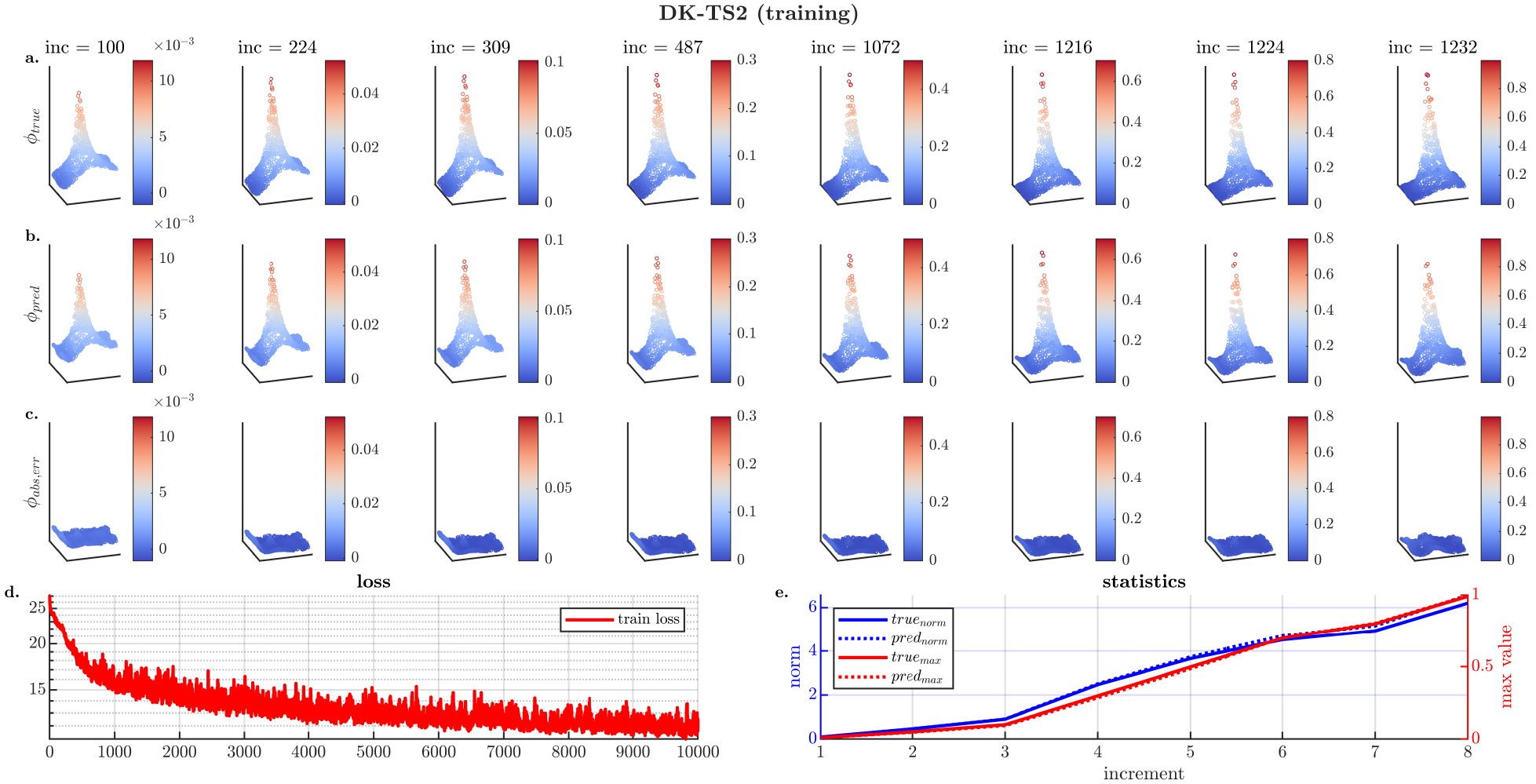}
    \caption{Training performance of DK-TS2. {\bf{a.}} True $\phi$ values. {\bf{b.}} Predicted $\phi$ values. {\bf{c.}} Absolute error $\phi$ values. {\bf{d.}} Evolution of training loss. {\bf{e.}} Metrics of $\phi$, including its $L^2$ norm and maximum value.}
    \label{Fig:Figure_DK_TS2_training}
\end{figure}

The results of the DK-TS1 and DK-TS2 training are depicted in Figs. \ref{Fig:Figure_DK_TS1_training} and \ref{Fig:Figure_DK_TS2_training} respectively. In both figures, the top row displays the true $\phi$ profiles, the middle row presents the corresponding network predictions, and the bottom row highlights the absolute error distributions. Additionally, the evolution of the training loss is provided on the bottom-left of each panel, while supplementary metrics, including the $L^2$ norm of the field and its maximum value, are displayed on the bottom-right. Overall, both networks exhibit strong agreement with the reference fields, a trend that is particularly pronounced for the DK-TS2 variant. This superior performance of DK-TS2 is rather anticipated, since in DK-TS1 near-zero values and vanishing fluxes are enforced at the sensors of Zone D based on an idealized assumption. Nonetheless, DK-TS1 is still capable of capturing this wide range of $H-\phi$ profiles with sufficient accuracy.

\begin{figure}[H]
    \centering
    \includegraphics[width=0.8\textwidth]{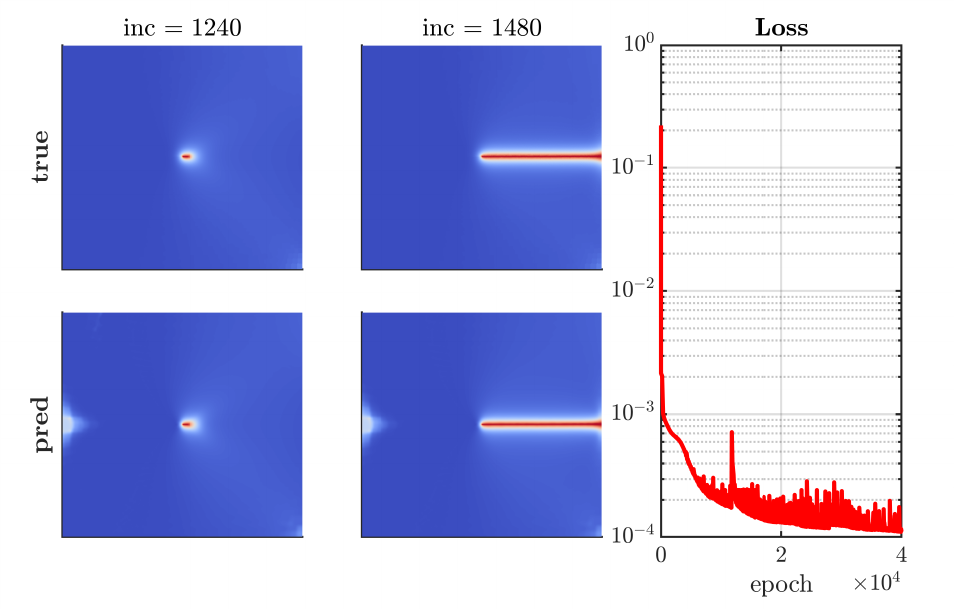}
    \caption{Training performance of the developed CNN. The true and predicted $\phi$ values for the two training increments are shown in the top and bottom row respectively, along with the evolution of the loss function.}
    \label{Fig:Figure_CNN_A16_training}
\end{figure}

Next, we develop the CNN model to capture the propagation stage of phase-field. The network consists of four consecutive convolution layers, with subsequent activation functions as presented in Section \ref{Sec:Method_PICNN}. The $H$ profile is first transformed into an image-like representation of size $600 \times 600$ pixels. The network is trained on two propagation increments, extracted from the FEM analysis, using the Adam optimizer with 40000 epochs and $10^{-3}$ learning rate. A comparison between the true and predicted $\phi$ values for the trained CNN is depicted in Fig. \ref{Fig:Figure_CNN_A16_training}, along with the evolution of the training loss.

\section{Numerical results}
\label{Sec:Numerical_results}

In this section we present the results of our numerical experiments. A list of the different cases is presented below:

\begin{itemize}

\item Case 1: Single-pass IFENN vs FEM on the training geometry, using the DK-TS2 for initiation.

\item Case 2: Multi-pass IFENN vs FEM on the training geometry, using the DK-TS2 for initiation.

\item Case 3: Single-pass IFENN vs FEM on the training geometry, using the DK-TS1 for initiation.

\item Case 4: Single-pass IFENN vs FEM on the testing geometry, using the DK-TS2 for initiation.

\end{itemize} 

\begin{figure}[H]
    \centering
    \includegraphics[width=0.8\textwidth]{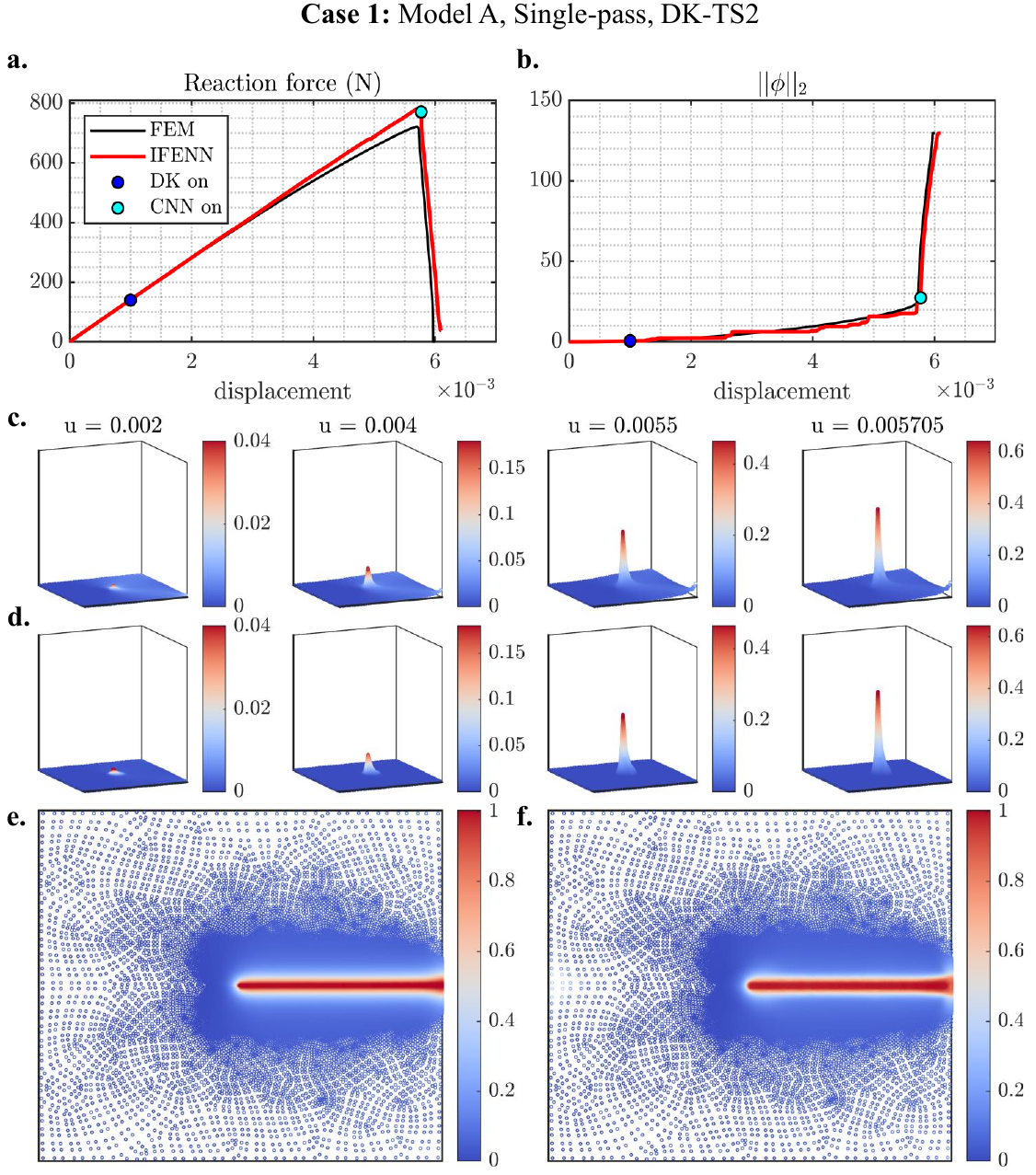}
    \caption{Numerical results: Case 1. {\bf{a.}} Reaction force-displacement curves. {\bf{b.}} Evolution of phase-field second norm. {\bf{c.}} 3D views of phase-field contours at various initiation increments from FEM (top) and IFENN (bottom).}
    \label{Fig:Figure_Case1}
\end{figure}

Fig. \ref{Fig:Figure_Case1}a and \ref{Fig:Figure_Case1}b show the evolution of the force-displacement and phase-field norm for the first example. The IFENN analysis begins with a fully coupled FEM solver until the $100^{th}$ increment, at which the physics-informed DeepOKAN is activated. We note that the network could have been readily activated from earlier increments, however it is intentionally initiated at the first increment that the network encountered during its training phase. Fig. \ref{Fig:Figure_Case1}c and \ref{Fig:Figure_Case1}d depict the phase-field profiles at different initiation increments between FEM and IFENN respectively, demonstrating excellent agreement between the two methods. Once the phase-field variable attains a maximum value of $0.90$, the physics-informed CNN is activated to capture the subsequent post-peak propagation stage. The phase-field contours at the end of the propagation stage are shown in Fig. \ref{Fig:Figure_Case1}e and \ref{Fig:Figure_Case1}f for FEM and IFENN respectively, again showing very good resemblance.

\begin{figure}[H]
    \centering
    \includegraphics[width=0.8\textwidth]{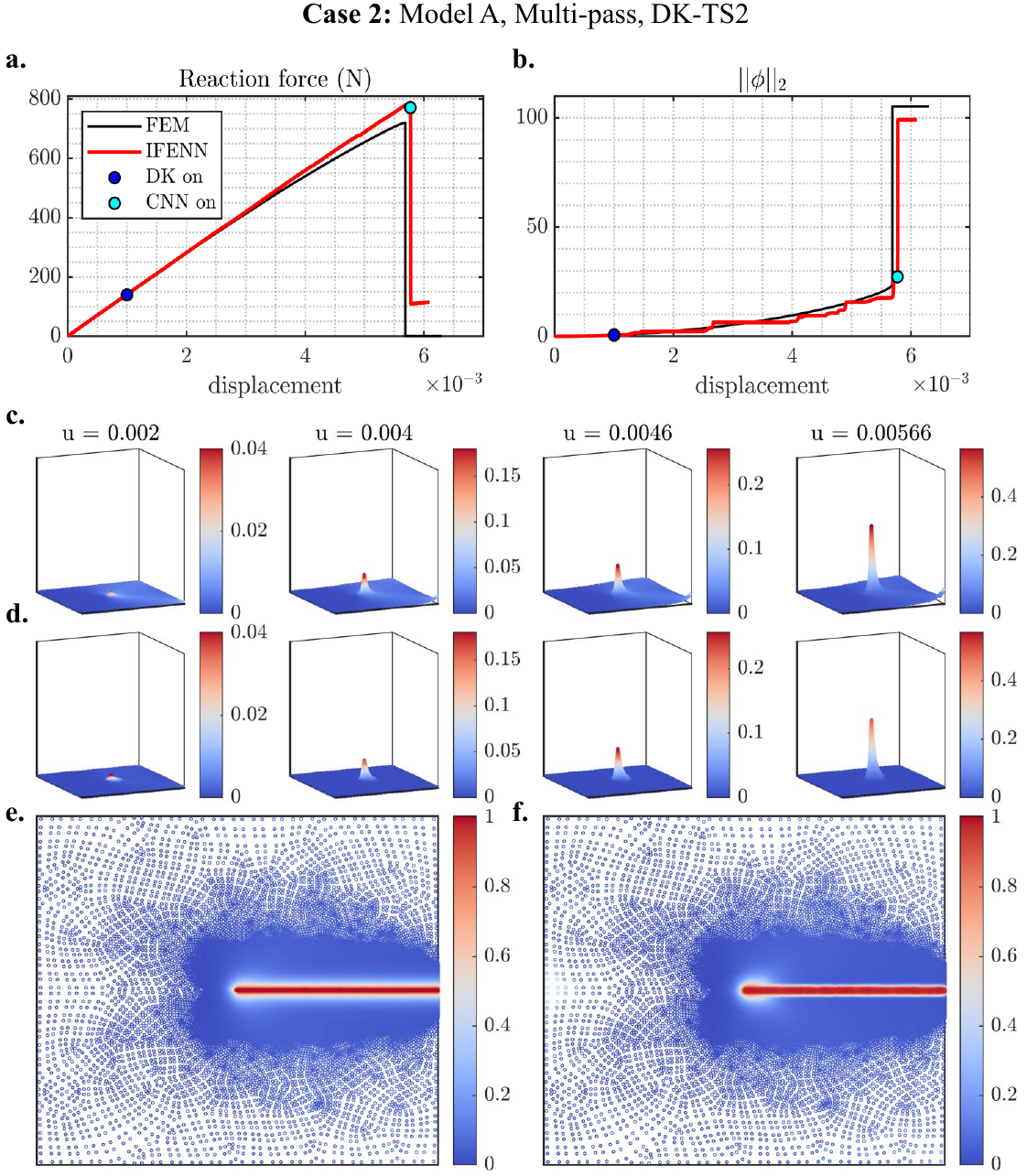}
    \caption{Numerical results: Case 2. {\bf{a.}} Reaction force-displacement curves. {\bf{b.}} Evolution of phase-field second norm. {\bf{c.}} 3D views of phase-field contours at various initiation increments from FEM (top) and IFENN (bottom).}
    \label{Fig:Figure_Case2}
\end{figure}

Next, we demonstrate the capability of the hybrid analysis to operate within a multi-pass staggered scheme, again on Model A. The numerical tolerance is set to $tol = 10^{-3}$, defining the threshold below which the displacement residual must fall to achieve convergence. The results of this investigation are shown in Fig. \ref{Fig:Figure_Case2}, following a layout identical to that of Fig. \ref{Fig:Figure_Case1}. The same criteria are established for the activation of the two networks, and an excellent match is obtained across all compared quantities between the FEM and IFENN frameworks. We emphasize that the force-displacement curve exhibits a nearly vertical drop, accompanied by a corresponding vertical increase in the phase-field norm. This behavior stems from the highly brittle nature of the crack propagation in the single-notch tension problem, which occurs almost instantaneously within a single load increment. Both methodologies successfully capture this brittle behavior, further highlighting the robustness of IFENN.

\begin{figure}[H]
    \centering
    \includegraphics[width=0.8\textwidth]{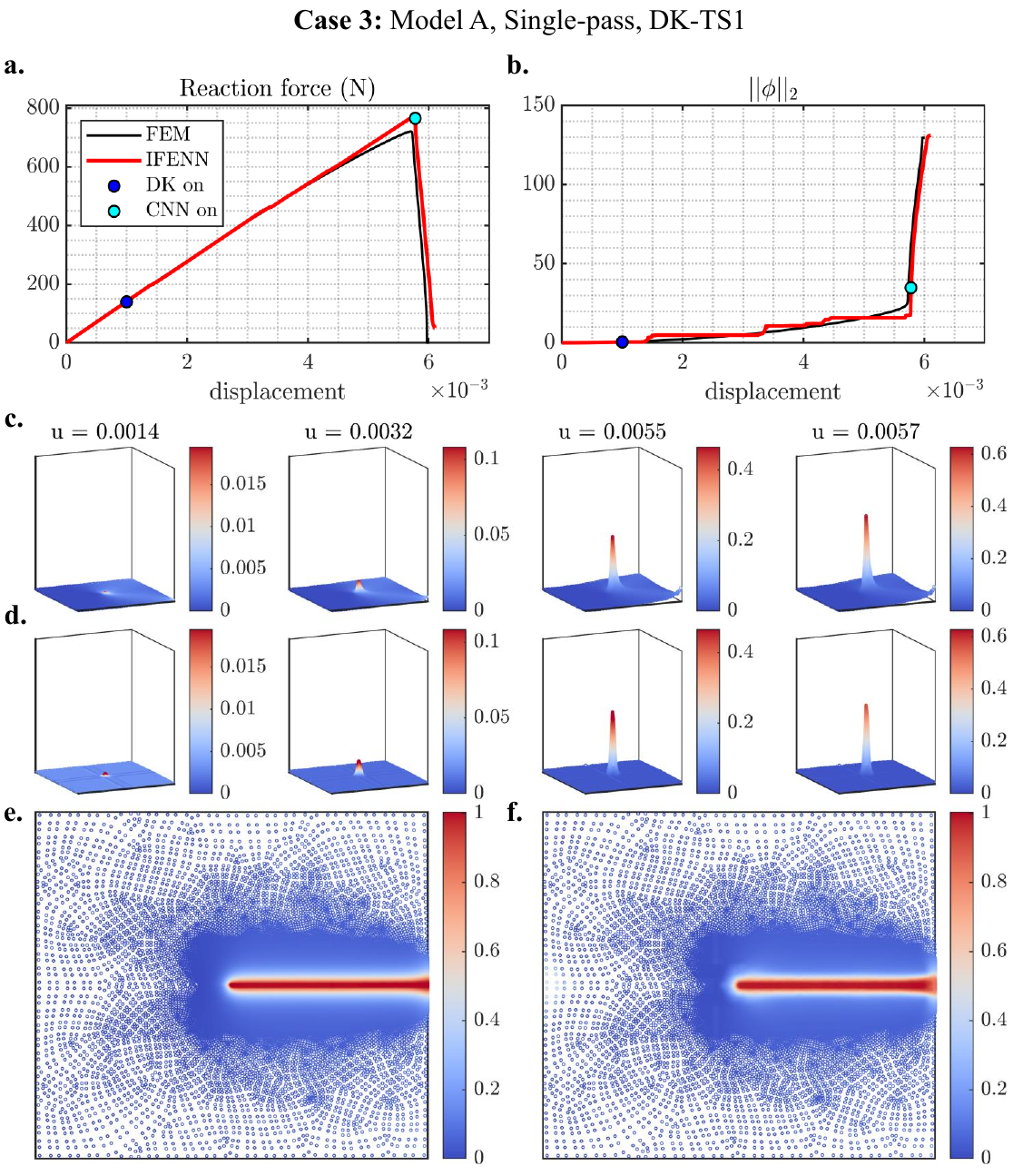}
    \caption{Numerical results: Case 3. {\bf{a.}} Reaction force-displacement curves. {\bf{b.}} Evolution of phase-field second norm. {\bf{c.}} 3D views of phase-field contours at various initiation increments from FEM (top) and IFENN (bottom).}
    \label{Fig:Figure_Case3}
\end{figure}

Having demonstrated the performance of IFENN under the weight-function approach, we now evaluate its implementation utilizing a DeepOKAN that is trained via the artificial boundary condition method (TS1). The results of this analysis are shown in Fig. \ref{Fig:Figure_Case3}. A similar layout of the compared variables is adopted as before, and an excellent agreement between the two methods is obtained, with the results of the hybrid analysis closely following the FEM-obtained trends during the entire evolution. Here we emphasize that the far-field predictions of the DeepOKAN are near-zero \textit{by default}, eliminating the need for additional 'filters' that vanish the field, as was done in the previous two cases. Therefore, this investigation provides clear evidence regarding the feasibility of applying artificial boundary conditions for generalizable modeling of phase-field initiation across arbitrary meshes.

\begin{figure}[H]
    \centering
    \includegraphics[width=0.8\textwidth]{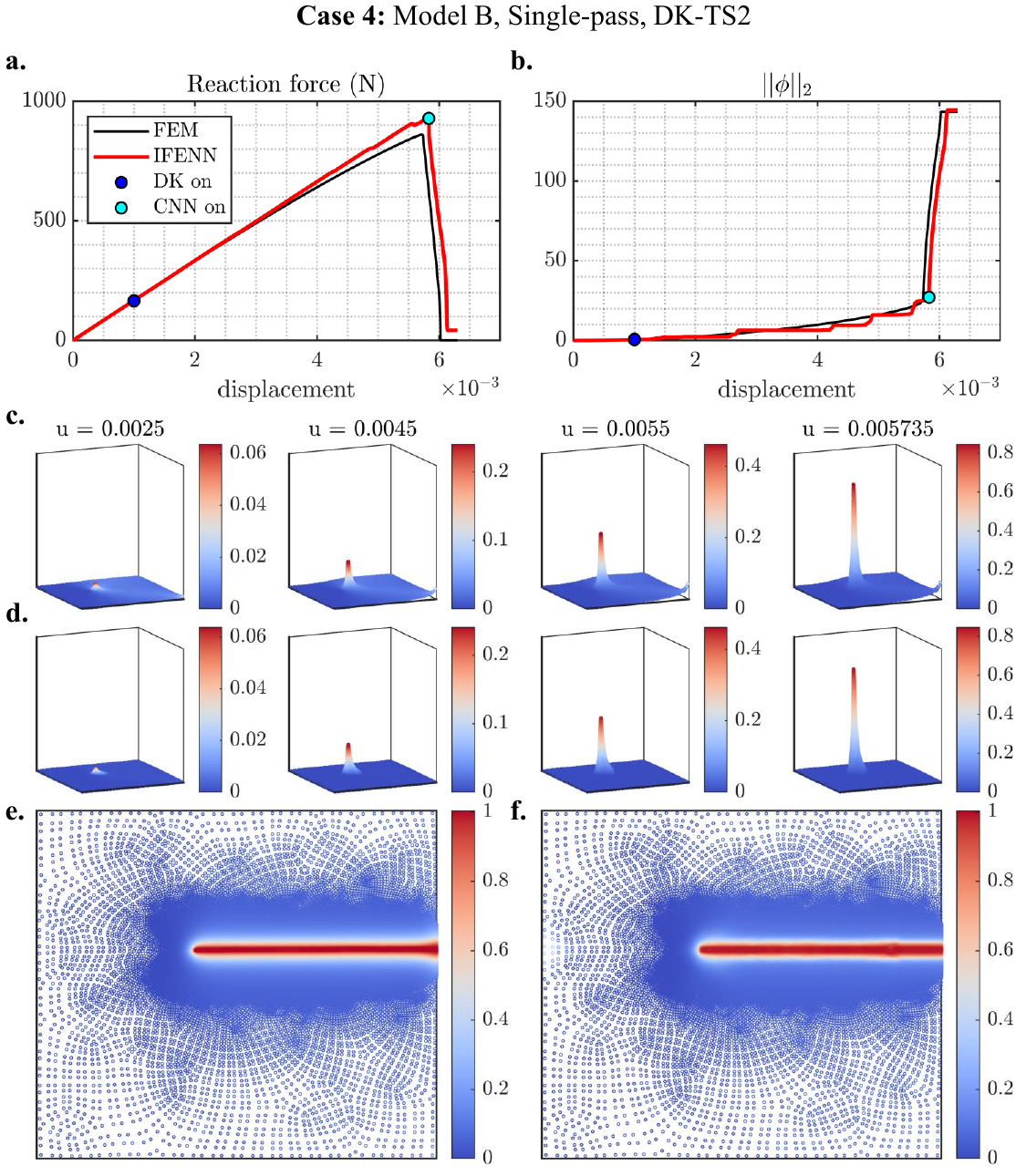}
    \caption{Numerical results: Case 4. {\bf{a.}} Reaction force-displacement curves. {\bf{b.}} Evolution of phase-field second norm. {\bf{c.}} 3D views of phase-field contours at various initiation increments from FEM (top) and IFENN (bottom).}
    \label{Fig:Figure_Case4}
\end{figure}

So far, all IFENN analyses have been conducted on Model A, which is the same geometry that was used for the training of both networks. In this final example, we extend the applicability of these pre-trained networks to a new geometry, designated as Model B in Fig. \ref{Fig:Figure_geometries}b, without requiring any re-training. Fig. \ref{Fig:Figure_Case4} illustrates the results of this investigation, utilizing the DeepOKAN that was trained with the weight-function method. A very close match between FEM and IFENN is observed again, with both the force-displacement and phase-field norm curves closely aligning with each other. The 3D views of phase-field at different initiation increments provide further evidence that the trained DeepOKAN captures correctly the extremely localized initiation phase. Overall, these findings demonstrate that the proposed physics-based hybrid framework possesses excellent generalization capabilities across different finite element meshes, without the need for expensive re-training (transfer learning, etc.).

\section{Summary, conclusions and outlook}
\label{Sec:Conclusions}

In this work, we present for the first time a complete IFENN framework for capturing the entire evolution of phase-field fracture in quasi-brittle materials. 
Building on prior developments of the IFENN pipeline, we present several methodological advancements that extend the framework's applicability in this challenging setup. The initiation stage is captured by a DeepOKAN, while a CNN is adopted for the propagation stage. Both networks operate on unstructured meshes and they are trained strictly on the physical relationship between the history variable and phase-field, omitting any labeled datasets from the training process. A total of just 10 samples is used for training, which decreases drastically the offline computational cost. To resolve the extreme localization of the initiation stage, we introduce two distinct training strategies. The first is based on a novel set of artificial boundary conditions, while the second adopts a local weight-function approach. Each strategy is accompanied by a corresponding sampling method for the input sensors, which effectively optimizes the trade-off between training cost and inference accuracy. Additionally, we propose a sensor location-based method that successfully generalizes the trained operator to unseen domain geometries.

The numerical results of the hybrid analysis demonstrate that IFENN matches benchmark FEM solutions with excellent agreement across both stages. This finding holds true regardless of the DeepOKAN training strategy. Additionally, the two networks that were trained once, on a benchmark geometry, were successfully implemented within IFENN against an unseen geometry which featured a different crack location and mesh density, yielding again very good agreement against FEM. We emphasize that no transfer learning or other form of re-training was used in the latter case, which evidently supports both the efficiency of the offline stage and the generalizability of the trained networks. 

Overall, this work establishes a comprehensive and physics-rooted hybrid framework that offers efficient and generalizable modeling of phase-field fracture. At the same time, several key challenges still persist. Future developments will focus on scaling this architecture to 3D domains, to demonstrate its relevance and scalability in real-world problems. Furthermore, ongoing research aims to expand IFENN's capability to capture more complex crack patterns, such as Mode II/III loading, crack branching, and crack coalescence. Finally, incorporating additional coupled phenomena such thermal and hydraulic cracking remains another critical direction, as the benefits of the hybrid solver are expected to be even more pronounced in this multi-physics context.

\section*{Acknowledgments}
\label{Section:Acknowledgments}

This work was partially supported by the Sand Hazards and Opportunities for Resilience, Energy, and Sustainability (SHORES) Center, funded by Tamkeen under the NYUAD Research Institute Award CG013. The authors would also like to acknowledge the support of the NYUAD Center for Research Computing for providing resources, services, and skilled personnel.

\section*{Data availability}
\label{Section:Data_Availability}

Data will be made available upon reasonable request.

\newpage
\bibliography{bibliography}

\end{document}